\newcommand{\ket}[1] { | {#1} \rangle }
\newcommand{\bra}[1] { \langle {#1} | }
\newcommand{\braket}[2] { \langle {#1} | {#2} \rangle}

\newcommand{\zeroKet}{\ket{0}}
\newcommand{\zeroBra}{\bra{0}}
\newcommand{\phiKet}{\ket{\phi}}
\newcommand{\psiKet}{\ket{\psi}}
\newcommand{\PsiKet}{\ket{\Psi}}
\newcommand{\PsiBra}{\bra{\Psi}}
\newcommand{\AKet}{\ket{A}}
\newcommand{\LKet}{\ket{L}}
\newcommand{\RKet}{\ket{R}}

\newcommand{\PsiPlusKet} {\PsiKet^{+}}
\newcommand{\PsiMinusKet} {\PsiKet^{-}}
\newcommand{\PsiPlusMinusKet} {\PsiKet^{\pm}}

\newcommand{\ourExamplePsi}
{
	\psiKet =
	\frac{1}{\sqrt{2}} ( c_{A\uparrow }^{\dag}+c_{B\uparrow }^{\dag} )    
	\frac{1}{\sqrt{2}} ( c_{A\downarrow }^{\dag}+c_{B\downarrow }^{\dag} )
	\zeroKet
}

\newcommand{\ourExampleRho}
{
	\frac{1}{4}
	( c_{A\uparrow }^{\dag}+c_{B\uparrow }^{\dag} )
	( c_{A\downarrow }^{\dag}+c_{B\downarrow }^{\dag} )
	\zeroKet \zeroBra
	( c_{A\downarrow }+c_{B\downarrow } )
	( c_{A\uparrow }+c_{B\uparrow } )
}

\newcommand{\upup}{\uparrow \uparrow}
\newcommand{\updown}{\uparrow \downarrow}
\newcommand{\downup}{\downarrow \uparrow}
\newcommand{\downdown}{\downarrow \downarrow}

\newcommand{\aup}{A \!\! \uparrow}
\newcommand{\adown}{A \!\! \downarrow}
\newcommand{\bup}{B \!\! \uparrow}
\newcommand{\bdown}{B \!\! \downarrow}

\newcommand{\aupup}{A \!\! \upup}
\newcommand{\aupdown}{A \!\! \updown}
\newcommand{\adownup}{A \!\! \downup}
\newcommand{\adowndown}{A \!\! \downdown}

\newcommand{\upone}{\uparrow\!\!(1)}
\newcommand{\uptwo}{\uparrow\!\!(2)}
\newcommand{\downone}{\downarrow\!\!(1)}
\newcommand{\downtwo}{\downarrow\!\!(2)}

\newcommand{\naup}{n_{A\uparrow}}
\newcommand{\nadown}{n_{A\downarrow}}
\newcommand{\nbup}{n_{B\uparrow}}
\newcommand{\nbdown}{n_{B\downarrow}}
\newcommand{\nbupprime}{n'_{B\uparrow}}
\newcommand{\nbdownprime}{n'_{B\downarrow}}

\newcommand{\aDiagonalElement}[1]
{\bra{\naup,\nadown} {#1} \ket{\naup,\nadown}}

\newcommand{\bMatrixElement}[1]
{\bra{\nbup,\nbdown} {#1} \ket{\nbupprime,\nbdownprime}}

\newcommand{\infinitesimalU}{(1-i \epsilon H)}
\newcommand{\oneSiteTwoParticleH}{U \naup \nadown}

\newcommand{\rhoHat}{\hat{\rho}}

\newcommand{\rhoB}{\rhoHat _{B}}

\newcommand{\siteBOccupationNoBasis}
{
	\{\nbup,\nbdown\} = \{0,0\}, \{1,0\}, \{0,1\}, \{1,1\}
}

\newcommand{\schliemannState}[1]
{
	\ket{w} = \sum _{a,b\in \{1,2,3,4\}}
		w_{ab} {#1}_{a}^{\dag} {#1}_{b}^{\dag} \zeroKet
}

\newcommand{\upUpKet}{\ket{\upup}}
\newcommand{\upDownKet}{\ket{\updown}}
\newcommand{\downUpKet}{\ket{\downup}}
\newcommand{\downDownKet}{\ket{\downdown}}

\newcommand{\ourExampleRhoB}
{
	\frac{1}{4}
	\begin{pmatrix}
		1 & 0 & 0 & 0 \\
		0 & 1 & 0 & 0 \\
		0 & 0 & 1 & 0 \\
		0 & 0 & 0 & 1
	\end{pmatrix}
}

\newcommand{\half}{\frac{1}{2}}
\newcommand{\quarter}{\frac{1}{4}}
\newcommand{\eighth}{\frac{1}{8}}

\newcommand{\oneOverRootTwo}{\frac{1}{\sqrt{2}}}

\newcommand{\woottersGeneralState}
{
	\phiKet = a\upUpKet + b\upDownKet + c\downUpKet + d\downDownKet,
	\nonumber \\
	|a|^{2} + |b|^{2} + |c|^{2} + |d|^{2} = 1
}

\newcommand{\siteAUpPlusSiteBUp}
{
	\frac{1}{\sqrt{2}}(A\uparrow + B\uparrow)
}

\newcommand{\siteADownPlusSiteBDown}
{
	\frac{1}{\sqrt{2}}(A\downarrow + B\downarrow)
}

\newcommand{\upDownSlaterDet}
{
	\begin{vmatrix}
		\upone & \downone \\
		\uptwo & \downtwo
	\end{vmatrix}
}

\newcommand{\downUpSlaterDet}
{
	\begin{vmatrix}
		\downone & \upone \\
		\downtwo & \uptwo
	\end{vmatrix}
}

\newcommand{\schliemannEtaDef}
{
	\eta (w) := | \braket{\widetilde{w}}{w} |
}

\newcommand{\schliemannWInnerProductDef}
{
	\braket{\widetilde{w}}{w} = &&
	\sum _{abcd} \widetilde{w}_{ab}^{*} w_{cd}
	\zeroBra c_{b} c_{a} c_{c}^{\dag} c_{d}^{\dag} \zeroKet
	\nonumber \\
	&& = \sum _{abcd} \epsilon ^{abcd} w_{ab} w_{cd}
	\nonumber \\
	&& = 8(w_{12} w_{34} + w_{13} w_{42} + w_{14} w_{23})
}

\newcommand{\schliemannWInnerProductDefModulus}
{
	| \braket{\widetilde{w}}{w} |
	&& =
	|\epsilon ^{abcd} w_{ab} w_{cd}|
	\nonumber \\	
	&& = |8(w_{12} w_{34} + w_{13} w_{42} + w_{14} w_{23})|
}

\newcommand{\modifiedHoppingH}
{
	H = t (
		c_{A\uparrow}^{\dag} c_{B\downarrow} +
		c_{B\downarrow}^{\dag} c_{A\uparrow}
	)
}

\newcommand{\iet}{i \epsilon t}

\newcommand{\occupationNoBasisForWoottersGeneralState}
{
	\ket{ n_\aup n_\adown n_\bup n_\bdown }
}

\newcommand{\woottersGeneralStateInOccupationNoBasis}
{
	a\ket{1010} + b\ket{1001} + c\ket{0110} + d\ket{0101}
}

\newcommand{\woottersRhoBInOccupationNoBasis}
{
	\rhoB =
	\begin{pmatrix}
		|a|^{2} + |c|^{2}			& ab^{*} + cd^{*} \\
		a^{*}b + c^{*}d				& |b|^{2} + |d|^{2}
	\end{pmatrix}
}

\newcommand{\woottersRhoBUnsimplifiedEigenvalues}
{
	\half(1-\sqrt{1-4|ad-bc|^{2}}),
	\half(1+\sqrt{1-4|ad-bc|^{2}})
}

\newcommand{\woottersTauDefn} {\tau = 4|ad-bc|^{2}}
\newcommand{\woottersXDefn} {x = \half(1+ \sqrt{1-\tau})}

\newcommand{\omarPlusPlusInputState}
{
	\oneOverRootTwo
	(
		a_{A1 \uparrow}^{+} a_{A2 \downarrow}^{+} \pm
		a_{A1 \downarrow}^{+} a_{A2 \uparrow}^{+}
	)
	\oneOverRootTwo
	(
		a_{B1 \uparrow}^{+} a_{B2 \downarrow}^{+} \pm
		a_{B1 \downarrow}^{+} a_{B2 \uparrow}^{+}
	)
	\nonumber \\
}

\newcommand{\rhoOneIn} {\rhoHat_{\text{1,in}}}
\newcommand{\rhoOneLeftIn} {\rhoHat_{\text{1L,in}}}
\newcommand{\rhoOneOut} {\rhoHat_{\text{1,out}}}
\newcommand{\rhoOneLeftOut} {\rhoHat_{\text{1L,out}}}

\newcommand{\omarInputRhoOne}
{
	\rhoOneIn =
	\begin{pmatrix}
		0 & . & .					& .					& .					& .					\\
		. & 0 & .					& .					& .					& .					\\
		. & . & \quarter	& .					& .					& .					\\
		. & . & .					& \quarter	& .					& .					\\
		. & . & .					& .					& \quarter	& .					\\
		. & . & .					& .					& .					& \quarter
	\end{pmatrix}
}

\newcommand{\omarInputRhoOneLeft}
{
	\rhoOneLeftIn =
	\begin{pmatrix}
		0 & .			& .			& .			\\
		. & 0			& .			& .			\\
		. & .			& \half	& .			\\
		. & .			& .			& \half
	\end{pmatrix}
}

\newcommand{\omarFermionOutputState}
{
	- \half \LKet_{1} \LKet_{2} - \half \RKet_{1} \RKet_{2}
	\nonumber \\
	- \half (\LKet_{1} \RKet_{2} + \RKet_{1} \LKet_{2})
	\nonumber \\
	+ \half
	(
		\ket{\aupdown}_{1} \ket{\adownup}_{2} +
		\ket{\adownup}_{1} \ket{\aupdown}_{2}
	)
	\nonumber \\
	- \half
	(
		\ket{\aupdown}_{1} \ket{\aupdown}_{2} +
		\ket{\adownup}_{1} \ket{\adownup}_{2}
	)
	\nonumber \\
	+
	(
		\ket{\aupup}_{1} \ket{\adowndown}_{2} +
		\ket{\adowndown}_{1} \ket{\aupup}_{2}
	)
}

\newcommand{\occupationNoRepresentationForOmar}
{
	\ket
	{
		n_{L1\uparrow} n_{L1\downarrow} n_{R1\uparrow} n_{R1\downarrow}
		n_{L2\uparrow} n_{L2\downarrow} n_{R2\uparrow} n_{R2\downarrow}
	}		
}

\newcommand{\omarOutputRhoOne}
{
	\rhoOneOut =
	\begin{pmatrix}
		\eighth	&	\eighth	& 0					& 0					& 0					& 0 \\
		\eighth	&	\eighth	& 0					& 0					& 0					& 0 \\
		0				&	0				& \eighth		& -\eighth	& 0					& 0 \\
		0				&	0				& -\eighth	& \eighth		& 0					& 0 \\
		0				& 0				& 0					& 0					& \quarter	& 0 \\
		0				& 0				& 0					& 0					& 0					& \quarter \\
	\end{pmatrix}
}

\newcommand{\omarOutputRhoOneLeft}
{
	\rhoOneLeftOut =
	\begin{pmatrix}
		\eighth		& .							& .							& . \\
		.					& \eighth				& .							& . \\
		.					& .							& \frac{3}{8}		& . \\
		. 				& .							& .							& \frac{3}{8}
	\end{pmatrix}
}

\newcommand{\omarRhoOneBasis}
{
	\ket
	{
		n_{L1\uparrow} n_{L1\downarrow} n_{R1\uparrow} n_{R1\downarrow}
	}
	\ =
	\{\ket{1100}, \ket{0011},
	\nonumber \\
	\ket{0110}, \ket{1001}, \ket{1010}, \ket{0101} \}
}

\newcommand{\omarRhoOneLeftBasis}
{
	\ket{n_{L1\uparrow} n_{L1\downarrow}}
	\ =
	\{ \ket{11}, \ket{00},\ket{01},\ket{10} \}
}

\newcommand{\antisymmetricCTwoKByCTwoKSpace}
{
	\mathcal{A} ( \mathcal{C}^{2K} \otimes \mathcal{C}^{2K} )
}

\newcommand{\schliemannSlaterDecomposition}
{
	\PsiKet =
	\frac{1}{\sqrt{\sum_{i=1}^{K} |z_{i}|^{2}}}
	\sum_{i=1}^{K}
	z_{i} f_{a1(i)}^{\dag} f_{a2(i)}^{\dag}  \zeroKet
}

\newcommand{\twoQubitSpinStatesToTwoParticleOccupationNosMapping}
{
	& \{ \sigma_{A} \sigma_{B} \} = &
	\{ \upUpKet,\upDownKet,\downUpKet,\downDownKet \}
	\nonumber \\
	\rightarrow &&
	\nonumber \\
	& \{ n_{1L \uparrow} n_{1L\downarrow} \} = &
	\{ \ket{00}, \ket{01},\ket{10},\ket{11} \}
}

\newcommand{\isomorphismForTwoQubitTeleportation}
{
	& \{ \sigma_{A} \sigma_{B} \} = &
	\{ \upUpKet,\upDownKet,\downUpKet,\downDownKet \}
	\nonumber \\
	\rightarrow &&
	\nonumber \\
	& \{ n_{A \uparrow} n_{A\downarrow} \} = &
	\{ \ket{00}, \ket{11},\ket{10},\ket{01} \}
}

\newcommand{\spinOnlyCNOTDefinition}
{
	\upUpKet \rightarrow \upDownKet,
	\upDownKet \rightarrow \upUpKet,
	\downUpKet \rightarrow \downUpKet,
	\downDownKet \rightarrow \downDownKet
}

\newcommand{\aliceTeleportationOccupationNoBasis}
{
	\{   \ket{ n_{C\uparrow} n_{C\downarrow} n_{A\uparrow} n_{A\downarrow} }  \}
	= &
	\{ \ket{1000},\ket{1010},\ket{1001},\ket{1011}, &
	\nonumber \\
	& \ket{0100},\ket{0110},\ket{0101},\ket{0111} \} &
	\nonumber \\
}

\newcommand{\fourMatrixNoughts}{0 & 0 & 0 & 0}

\newcommand{\cnotOnVirtualFirstQubitInDelocalizedState}
{
	\hat{U}_{\text{1st qubit}} =
	\begin{pmatrix}
		0 & 1 & 0 & 0 & \fourMatrixNoughts \\
		1 & 0 & 0 & 0 & \fourMatrixNoughts \\
		0 & 0 & 0 & 1 & \fourMatrixNoughts \\
		0 & 0 & 1 & 0 & \fourMatrixNoughts \\
		\fourMatrixNoughts & \fourMatrixNoughts \\
		\fourMatrixNoughts & \fourMatrixNoughts \\
		\fourMatrixNoughts & \fourMatrixNoughts \\
		\fourMatrixNoughts & \fourMatrixNoughts
	\end{pmatrix}
}

\newcommand{\actionOfVirtualFirstQubitCNOT}
{
	\ket{1000} \leftrightarrow \ket{1010},
	\ket{1001} \leftrightarrow \ket{1011},
	\nonumber \\
	\ket{01 n_{A\uparrow} n_{A\downarrow} } \ \text{unchanged}
}

\newcommand{\hamiltonianForCNOTOnVirtualFirstQubitInDelocalizedState}
{
	\hat{H} _{\text{1st qubit}} &=&
	\ket{10}_{C C} \bra{10}
	\biggl(
		\ket{00}_{A A} \bra{10} +
		\nonumber \\
		&& \quad \ket{10}_{A A} \bra{00} +
			\ket{11}_{A A} \bra{01} + \ket{01}_{A A} \bra{11}
	\biggr)
	\nonumber \\
	&=&
	\half (\sigma_{z,C} + 1)
	\biggl(
		c_{A\uparrow}^{\dag} + c_{A\uparrow}
	\biggr)
}

\newcommand{\hamiltonianForCNOTOnVirtualFirstQubitInDelocalizedStateWithParticleSourceSink}
{
	\hat{H} _{\text{1st qubit}} &=&
	\ket{10}_{C C} \bra{10}
	\biggl(
		c_{A\uparrow}^{\dag} c_{D} + c_{D}^{\dag} c_{A\uparrow}
	\biggr)
}

\newcommand{\hamiltonianForCNOTOnVirtualSecondQubitInDelocalizedState}
{
	\hat{H} _{\text{2nd qubit}} &=&
	\half (\sigma_{z,C} + 1)
	\biggl(
		c_{A\uparrow}^{\dag} c_{A\downarrow}^{\dag} +
		c_{A\downarrow} c_{A\uparrow}
	\biggr)
}

\newcommand{\basesForHamiltonianForCNOTOnVirtualFirstQubitInDelocalizedState}
{
	\ket{ n_{C\uparrow} n_{C\downarrow} }
	\text{\  and \ }
	\ket{ n_{A\uparrow} n_{A\downarrow} }
}

\newcommand{\coherentStateDefinition}
{
	\ket{\alpha}_{D} =	\text{exp} (-\frac{1}{2} |\alpha|^{2})
				\sum_{n}
				\frac{\alpha^{n}}{(n!)^{1/2}} \ket{n}_{D}
}

\documentclass[aps,twocolumn,showpacs,preprintnumbers,amsmath,amssymb]{revtex4}
\usepackage{graphicx}% Include figure files
\usepackage{dcolumn}% Align table columns on decimal point
\usepackage{bm}% bold math

\begin{document}

\title
{
	Describing mixed spin-space entanglement of \\
	pure states of indistinguishable particles \\
	using an occupation number basis
}

\author{J. R. Gittings}
	\email{joe.gittings@ucl.ac.uk}
	\homepage{http://www.cmmp.ucl.ac.uk/~jrg/}
\author{A. J. Fisher}
	\email{andrew.fisher@ucl.ac.uk}
	\homepage{http://www.cmmp.ucl.ac.uk/~ajf/}
\affiliation{Department of Physics and Astronomy, University College London, Gower St, London WC1E 6BT, UK}

\date{\today}

%=================================================================%
\begin{abstract}

Quantum mechanical entanglement is a resource for quantum computation, quantum teleportation, and quantum cryptography. The ability to quantify this resource correctly has thus become of great interest to those working in the field of quantum information theory. In this paper, we show that all existing entanglement measures but one fail important tests of fitness when applied to $n$ particle, $m$ site states of indistinguishable particles, where $n,m \ge 2$. The accepted method of measuring the entanglement of a bipartite system of distinguishable particles is to use the von Neumann entropy of the reduced density matrix of one half of the system. We show that expressing the full density matrix using a site-spin occupation number basis, and reducing with respect to that basis, gives an entanglement which meets all currently known fitness criteria for systems composed of either distinguishable or indistinguishable particles.

We consider an output state from a previously published thought experiment, a state which is entangled in both spin and spatial degrees of freedom, and show that the site entropy measure gives the correct total entanglement. We also show how the spin-space entanglement transfer occurring within the apparatus can be understood in terms of the transfer of probability from single-occupancy to double-occupancy sectors of the density matrix.

\end{abstract}

%=================================================================%
% PACS, the Physics and Astronomy Classification Scheme.

\maketitle

%=================================================================%
\section{Introduction}

The peculiarly non-local correlations exhibited by the states of quantum systems are key to the implementation of quantum information processing technologies, such as quantum computation and quantum teleportation. However, it is easily shown that the correlations due to the (anti)symmetrization of the states of indistinguishable bosons (fermions) are not themselves a physically useful resource for quantum information technologies: for example, there is no measurement we can make locally on a fermion in a localized state which is affected by the existence of identical fermions in other parts of the universe \cite{peresBook}. However, it is possible to produce entanglement that is a resource for QIT by suitable preparation: for example, by producing a $\PsiKet^{-}$ Bell state of the spins of two fermions. Indeed, in practice, many potential implementations of QIT involve identical particles (such as photons, electrons, or protons) as `carriers' of entanglement. It is therefore important to be able to quantify the degree of `useful' entanglement in a system of identical particles.

Discussion of the entanglement between pure states of indistinguishable particles has previously been dealt with almost as a separate topic from that of distinguishable particles. It is the aim of this paper to show that the entanglement of pure states of either type of particle can be described within the same theoretical framework. This framework involves the von Neumann entropy of the reduced density matrix for the subsystem whose entanglement with the rest of the system we wish to find, expressed in an occupation number basis \cite{zanardi2001}. This also allows us to understand better the division between spin and spatial entanglement in systems where both may exist, and the manner in which entanglement may be transferred between spin and space. It is important to emphasize that we consider in this paper only pure states of the full system. It is already known that for such states the von Neumann entropy provides the correct measure of entanglement between two distinguishable subsystems \cite{preskillBook}. We do not address the case of an overall mixed state, for which the definition of an entanglement measure is more subtle \cite{henderson2000}.

In section \ref{section:partitioning} we discuss the partitioning of Hilbert space that is implicit to any meaningful definition of entanglement. In section \ref{section:measureReview} we review some requirements for a successful entanglement measure, and consider the extent to which three potential definitions meet these requirements. In section \ref{section:siteEntropyMeasure} we show that Zanardi's site entropy measure passes all the tests, and can be related to the conventional definition of entanglement in the limit where the exchange symmetry of the particles is irrelevant. Finally, in section \ref{section:entanglementTransfer} we use Zanardi's measure to discuss spin-space entanglement transfer.

%=================================================================%
\section{Methods of partitioning Hilbert space of two entangled spinful particles}
\label{section:partitioning}

Implicit to any measure which attempts to describe the entanglement of two subsystems is an assumption about the correct manner in which to partition the total Hilbert space. In this section we consider the requirements for a correct partitioning, and look at how this is actually performed by existing entanglement measures. We will frequently need to talk about the states of internal degrees of freedom of particles. Therefore, for brevity we will henceforth refer to any states of such internal degrees of freedom simply as `spin states'.

%.................................................................%
\subsubsection{Requirements for partitioning}

\paragraph*{Tensor product structure}

In order to express entanglement between two components of an entangled system, some kind of partitioning of their Hilbert space is necessary in order to identify the `components'. Our aim is to quantify the entanglement resource shared between parts $A$ and $B$ of a composite quantum system. These parts may be identified with particles (in the case of a state of the system where the particles are localized), with sites (in the case of a state of the system where the particles are delocalized over sites), or with some arbitrary subdivision of an experimental apparatus (as we will examine in section \ref{omarSubsection}). For the purposes of the greater part of this paper, we will consider subsystems of a system as being synomymous with sites. But it is important to emphasize that our conclusions are more general: they apply to any division of a system into subsystems.

For entangled states of distinguishable particles (or particles which are effectively distinguishable because of their localization) we would normally use a tensor product structure
$\mathcal{H} = \mathcal{H}_{A} \otimes \mathcal{H}_{B}$
where
$\mathcal{H}_{A}$ and $\mathcal{H}_{B}$ are Hilbert spaces for states of particles in parts $A$ and $B$. It is important that we correctly partition the Hilbert space because this ensures that basic operations such as the partial trace $\rhoB = \text{tr}_{B} \rho$ are valid. The partial trace is the correct and only way to describe the properties of one part of a composite quantum system when nothing is known about the other parts, as it gives the correct measurement statistics for observations on that subsystem
\cite{nielsenAndChuang}.

But if we try to use the tensor product structure partitioning for entangled states of indistinguishable particles, we run into two problems:
\begin{itemize}
	\item The Hilbert space of two indistinguishable particles is a
	symmetric or antisymmetric product, not a direct product.
	\item There is no correspondence between the particles and
	the subsystems used in the partitioning.
\end{itemize}

\paragraph*{Delocalization}

For spin-only entangled states of distinguishable particles---i.e. states where we have unambigously given one particle to Alice, and the other to Bob---the phrase `the states of Alice's spin' is completely equivalent to the phrase `the states of Alice's particle'. There is no ambiguity about which particle Alice has in her possession at any time, and therefore there is no logical difference between a one-site (local) unitary transformation, and a one-particle (possibly non-local) unitary transformation. Thus when deciding on a basis in which to describe the spin-only entanglement of a system of distinguishable particles it may seem a matter of taste whether spin states should be assigned to particles, or to sites.

However, it is perfectly possible to write down states in which each particle is shared between Alice and Bob. An example of such a `spin-space entangled state' is obtained if we put particle 1 into $\siteAUpPlusSiteBUp$ and particle 2 into $\siteADownPlusSiteBDown$, where $A$, $B$ are site labels.

\paragraph*{Indistinguishability}

When the entangled particles are indistinguishable, we can no longer be sure which particle Alice has in her possession. The distinction between one-particle unitary transformations, and one-site unitary transformations becomes relevant. Entanglement should be invariant under one-site unitary transformations, but not necessarily under one-particle unitary transformations, which may generate entanglement if they involve both subsystems. An entanglement measure which works successfully for indistinguishable particles must respect this distinction.

The natural way to achieve this distinction is to use a basis
which assigns spin states to \textbf{sites} rather than \textbf{particles}.

%.................................................................%
\subsubsection{Partitioning used by existing entanglement measures}

When partitioning the total Hilbert space of two entangled quantum systems, we need to ask ourselves:
\begin{itemize}
	\item For indistinguishable subsystems: to what extent can my system be regarded as a symmetric/antisymmetric product of the single-subsystem states?
	\item For distinguishable subsystems: to what extent can my system be regarded as a direct product of the single-subsystem states?
\end{itemize}

In most descriptions of entanglement, the tensor product structure is used, for example in the entanglement measure introduced by Wootters \cite{woottersRoyalSoc1998}. This measure is suitable for spin-only entanglement of localized distinguishable particles. However, it does not describe which site a particle occupies, so is not suited to describing either entangled indistinguishable particles, or entangled states of distinguishable particles where the `particle' and `subsystem' divisions do not coincide.

One example where indistinguishable particles have been treated is by Schliemann \emph{et al.} \cite{schliemannQuantPh2001} \cite{schliemannCondMat2001}, who explicitly consider the antisymmetric product space belonging to two fermions, each of which inhabits a four-dimensional one-particle space. They write a general state in the six-dimensional two-particle Hilbert space as
\begin{eqnarray}
	\schliemannState{c}
\end{eqnarray}
where a,b run over the orthonormalized single particle states, and Pauli exclusion requires that the $4 \times 4$ coefficient matrix $w$ is antisymmetric: $w_{ab} = -w_{ba}$.

It may seem that Schliemann's partitioning is indeed in terms of sites rather than particles, since the single particle states are labelled by sites. But, as we shall see later, Schliemann's measure is derived by considering the number of elemental Slater determinants needed to expand the entangled state. It is therefore actually a particle-based, rather than a site-based, description of entanglement. As a consequence, as will be shown later in this paper, it suffers from a number of serious flaws; in particular, it is possible to devise one-site (i.e. local) transformations which generate `entanglement' according to the Schliemann measure.

%=================================================================%
\section{Review of existing entanglement measures}
\label{section:measureReview}

%-----------------------------------------------------------------%
\subsection{Desirable properties of any entanglement measure}

What are the desirable properties of an entanglement measure?

%.................................................................%
\paragraph*{Invariance under local unitary transformations.}
If a measure is correct, it should not be possible to generate `entanglement' using only unitary transformations local to a particular site.

%.................................................................%
\paragraph*{Non-invariance under non-local unitary transformations.}
Conversely, it should be possible to find non-local (i.e. multisite) unitary transformations which change the entanglement.

%.................................................................%
% Source for this is logbook 1 p92 (fermions) and logbook 2 p70 (bosons)
\paragraph*{Correct behaviour as distinguishability of subsystems A and B is lost.}

A correct measure should reflect the fact that entanglement is affected when the distinguishability of the subsystems involved is lost. A simple example of this is as follows. For two fermions whose spin degrees of freedom are maximally entangled, we require that as the overlap of the single-particle spatial wavefunctions approaches unity the entanglement should asymptotically approach zero. This is easily seen by considering the full expression for the Bell basis states in terms of Slater determinants.

If the two fermions are localized, one in site $A$ and one in site $B$, then the $\PsiPlusMinusKet$ Bell state can be written
\begin{eqnarray}
	\PsiPlusMinusKet \equiv \oneOverRootTwo(\upDownKet \pm \downUpKet)
\end{eqnarray}
where the full expression for $\upDownKet$ is
\begin{eqnarray}
	\upDownKet = \oneOverRootTwo
	\begin{vmatrix}
		\phi_{A}(1)\uparrow(1)	& \phi_{B}(1)\downarrow(1) \\
		\phi_{A}(2)\uparrow(2)	& \phi_{B}(2)\downarrow(2)
	\end{vmatrix}
\end{eqnarray}

When the two fermions are brought together to occupy the same site, the spatial parts of the two single-particle states coincide, i.e. $\phi_{A} \rightarrow \phi_{B}$, and we have
\begin{eqnarray}
	\upDownKet \rightarrow \oneOverRootTwo
	\begin{vmatrix}
		\phi(1)\uparrow(1)	&	\phi(1)\downarrow(1) \\
		\phi(2)\uparrow(2)	& \phi(2)\downarrow(2)
	\end{vmatrix}
	\nonumber \\
	= \frac {\phi(1) \phi(2)} {\sqrt{2}} \upDownSlaterDet
\end{eqnarray}
where $\phi$ is the \textbf{same} spatial state for sites $A$ and $B$.

A similar result is obtained for $\downUpKet$, but with an exchange of columns and therefore the same result applies for it as for $\upDownKet$ but with an overall minus sign. Hence $\upDownKet$ and $\downUpKet$ are now linearly dependent, and the behaviour of the entangled Bell state is:
\begin{eqnarray}
	\PsiPlusMinusKet
	\ \stackrel{\phi_{A} \rightarrow \phi_{B}}{\longrightarrow}
	&&
	\oneOverRootTwo
	\frac {\phi(1) \phi(2)} {\sqrt{2}}
	\nonumber \\
	&&
	\biggl(
		\upDownSlaterDet \pm \downUpSlaterDet
	\biggr)
	\nonumber \\
	= &&
	\half \  \phi(1) \phi(2)
	\biggl(
		\upone \! \downtwo - \downone \! \uptwo
		\nonumber \\ &&
		\pm ( \downone \! \uptwo - \upone \! \downtwo )
	\biggr)
\end{eqnarray}
and hence
\begin{eqnarray}
	\PsiMinusKet \rightarrow \phi(1)\phi(2) \upDownSlaterDet
\end{eqnarray}
up to a normalization factor, whereas
$\PsiPlusKet \rightarrow 0$ because of Pauli exclusion.

Thus the one ebit of entanglement present in a $\PsiPlusMinusKet$ state should be destroyed as the spatial overlap of the two fermions' wavefunctions asymptotically approaches unity---in the case of $\PsiPlusKet$ because the state itself is destroyed, and in the case of $\PsiMinusKet$ because the entangled Bell state becomes a non-entangled product state. (At least, this is the case if neither Alice nor Bob can measure with spatial resolution sufficient to determine the substructure of the spatial state $\phi$.) A correct entanglement measure should reflect this fact.

For a pair of bosons in the $\PsiPlusMinusKet$ state, exactly the same loss of entanglement would occur, although the behaviours of $\PsiPlusKet$ and $\PsiMinusKet$ are exchanged, due to the change of sign introduced by the use of permanents rather than determinants.

\newpage

%-----------------------------------------------------------------%
\subsection{Wootters measure for distinguishable particles \\ (tangle)}

Wootters
\cite{woottersRoyalSoc1998,hill1997,woottersPhysRevLett1998}
considers a particular state of two distinguishable particles:
\begin{eqnarray}
	\label{eqnWoottersGeneralState}
	\woottersGeneralState
\end{eqnarray}
where it is implicit that each particle occupies a definite spatial state. Since the tensor product decomposition of $H$ allows us to define a reduced density matrix $\rhoB$ describing the mixed state of system $B$, the von Neumann entropy of $\rhoB$ is a natural measure of entanglement. The Wootters entanglement is simply a reexpression of the von Neumann entropy of $\rhoB$, and is defined as
\begin{eqnarray}
	\label{eqnWoottersE}
	E = h[\frac{1}{2} (1 + \sqrt{1- \tau} )],
\end{eqnarray}
where
\begin{eqnarray}
	h(x)=-(x \text{log} _{2} x + (1-x) \text{log} _{2} (1-x))
\end{eqnarray}
and the quantity $\tau$ is known as the `tangle' and is defined by
\begin{eqnarray}
	\woottersTauDefn.
\end{eqnarray}	
Since $E_{\text{Wootters}}$ expresses the entropy of a single site, there is no single-site operation which can affect it.

The Wootters measure applies only when the particles are totally distinguishable by virtue of occupying distinct sites. But our aim is to describe more general states in which each particle occupies a superposition of sites---what happens if we simply go ahead and use the Wootters measure regardless? Since the Wootters measure does not depend on the nature of the spatial states, there is no way its value can change. So for example, there is no way that the Wootters entanglement of a $\PsiPlusMinusKet$ Bell state will ever be affected by the spatial overlap of the single-particle wavefunctions of the constituent particles.

%-----------------------------------------------------------------%
\subsection{Schliemann measure for fermions}

Schliemann \emph{et al.}
\cite{schliemannCondMat2001}
define the entanglement of spin states of a pair of fermions by
\begin{eqnarray}
	\schliemannEtaDef,
\end{eqnarray}
where the dual $\widetilde{w}$ of $w$ is defined by
\begin{eqnarray}
	\widetilde{w}_{ab} =
	\frac{1}{2} \epsilon ^{abcd} \overline{w}_{cd}
\end{eqnarray}
and the inner product is expressed as
\begin{eqnarray}
	\schliemannWInnerProductDef
	.
\end{eqnarray}
A similar definition was introduced for a pair of bosons by Pa\v{s}kauskas and You \cite{paskauskas2001}.

%.................................................................%
\paragraph*{Slater decomposition form.}
It is possible to relate the Schliemann measure $\eta$ to the number of elementary Slater determinants that are required to construct the entangled state. The Hilbert space for a two-fermion, $K$-site system is the antisymmetric space $\antisymmetricCTwoKByCTwoKSpace$. Any vector in this space can be represented in terms of single particle functions $f_{a(i)}^{\dag}$, which are members of the single-particle space $\mathcal{C}^{2K}$, by the Slater decomposition
\begin{eqnarray}
	\schliemannSlaterDecomposition
\end{eqnarray}
The number of non-zero coefficients $z_{i}$ required to construct $\PsiKet$, i.e. the number of elementary Slater determinants, is known as the Slater rank of the entangled state. Then for a two-fermion two-site system, $\PsiKet$ has Slater rank 1 (consists of a single Slater determinant) iff $\eta(\PsiKet) = 0$.

\begin{figure*}
	\caption
	{
		\label{fig:schliemann_entanglement}
		Schliemann entanglement $\eta$ of all Bell states vs overlap $S=| \braket{\phi_{a}}{\phi_{b}} |$ of single particle states.
	}

	\includegraphics{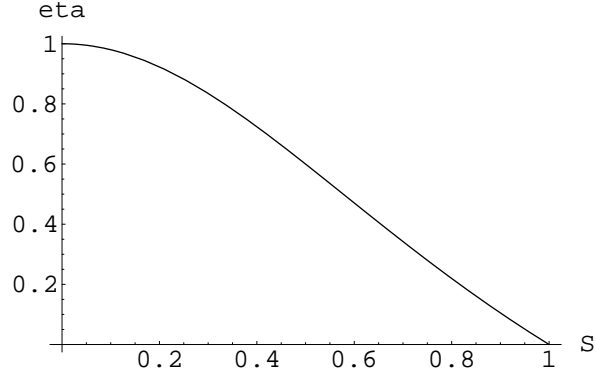}
\end{figure*}

%.................................................................%
\paragraph*{Behaviour as overlap of single particle wavefunctions is increased.}
This entanglement measure behaves correctly as the overlap is increased between the single-particle wavefunctions of the particles, as is shown in Figure \ref{fig:schliemann_entanglement}.

%.................................................................%
\paragraph*{Relation to Wootters measure.}
Let us consider how the Schliemann measure works for the class of states considered by Wootters:
\begin{eqnarray}
	\woottersGeneralState
\end{eqnarray}
In the representation used by Schliemann, we can write the $w$-matrices for the two-particle basis states in the
$\aup, \adown, \bup, \bdown$ basis as
\begin{eqnarray}
	w_{\upup} =
	\begin{pmatrix}
		0 	& 0 & \half	& 0 \\
		0 	& 0 & 0	& 0 \\
		-\half	& 0 & 0 & 0 \\
		0	& 0 & 0 & 0
	\end{pmatrix}
	,
	w_{\downdown} =
	\begin{pmatrix}
		0 & 0 		& 0 	& 0 \\
		0 & 0 		& 0	& \half \\
		0 & 0		& 0 	& 0 \\
		0 & -\half	& 0 	& 0
	\end{pmatrix}
	,
	&& \nonumber \\
	w_{\updown} =
	\begin{pmatrix}
		0 	& 0 & 0	& \half \\
		0 	& 0 & 0	& 0 \\
		0 	& 0 & 0 & 0 \\
		-\half	& 0 & 0 & 0
	\end{pmatrix}
	,
	w_{\downup} =
	\begin{pmatrix}
		0 & 0 		& 0 	& 0 \\
		0 & 0 		& \half	& 0 \\
		0 & -\half 	& 0 	& 0 \\
		0 & 0 & 0 & 0
	\end{pmatrix}
	.
	&& \nonumber \\	
\end{eqnarray}

Therefore the state considered by Wootters, \\ $\woottersGeneralState$, has the coefficient matrix
\begin{eqnarray}
	w = \half
	\begin{pmatrix}
		0	& 0	& a 	& b \\
		0	& 0	& c	& d \\
		-a	& -c	& 0	& 0 \\
		-b	& -d	& 0 	& 0
	\end{pmatrix}
\end{eqnarray}
and thus we obtain the relation
\begin{eqnarray}
	\schliemannWInnerProductDefModulus \nonumber \\
	&& = 2|ad-bc| = \sqrt{\tau}.
\end{eqnarray}

Hence for the state of two distinguishable particles considered by Wootters, the Schliemann measure $\eta$ is related to the Wootters `tangle' $\tau$ by
\begin{eqnarray}
	\eta = \sqrt{\tau}.
\end{eqnarray}

%.................................................................%
\paragraph*{Non-invariance under local unitary transformations.}
We can however easily show that there are local (one-site) unitary transformations that generate `entanglement' by the Schliemann measure. Consider this two particle state:
\begin{eqnarray}
	\label{eqn:ourExamplePsi}
	\ourExamplePsi.
\end{eqnarray}
The physical interpretation of this state is that it describes a doubly filled `molecular orbital'
\begin{eqnarray}
	\frac{\ket{A} + \ket{B}}{\sqrt{2}}
\end{eqnarray}
where $\ket{A},\ket{B}$ are the spatial states for sites $A$, $B$ respectively.

Its antisymmetric coefficient matrix is
\begin{eqnarray}
	w =
	\half \oneOverRootTwo \oneOverRootTwo
	\begin{pmatrix}
		.		& 1		& .		& 1 \\
		-1	& .		& -1	& . \\
		.		& 1		& .		& 1 \\
		-1	& .		& -1	& .
	\end{pmatrix}
\end{eqnarray}
giving a Schliemann entanglement of $\eta = 0$ (no entanglement, since it is a single Slater determinant).

Now consider applying the infinitesimal one-site two-particle unitary transformation $\infinitesimalU$ with $H=\oneSiteTwoParticleH$. This purely local operation transforms the $w$-matrix to
\begin{eqnarray}
	w \rightarrow \quarter
		\begin{pmatrix}
		.									& 1-i \epsilon U		& .		& 1 \\
		-1+i \epsilon U		& .									& -1	& . \\
		.									& 1									& .		& 1 \\
		-1								& .									& -1	& .
		\end{pmatrix}
\end{eqnarray}
which gives a Schliemann entanglement of
$\eta = 8|\frac{-i \epsilon U}{4}| = 2 \epsilon U$
which is non-zero to first order in $\epsilon$. We have succeeded in generating Schliemann `entanglement' via a purely local unitary operation, something that it should not be possible to achieve.

%.................................................................%
\paragraph*{Invariance under non-local unitary transformations.}
Now consider an infinitesimal two-site one-particle unitary transformation. We would expect such a transformation to lead to a change of entanglement, yet we can construct an example under which the Schliemann measure is invariant. Our example is generated by a Hamiltonian describing intersite hopping accompanied by a spin-flip:
\begin{eqnarray}
	\label{eqnModifiedHoppingH}
	\modifiedHoppingH.
\end{eqnarray}
(The spin-flip is introduced so that our state is not an eigenvector of $H$). The Hamiltonian's action on our example state is
\begin{eqnarray}
	\label{eqnActionOfModifiedHoppingHOnOurExampleState}
	H \psiKet =
		- \frac{t}{2} c_{A\uparrow}^{\dag} c_{B\uparrow}^{\dag} \zeroKet
		+ \frac{t}{2} c_{B\downarrow}^{\dag} c_{A\downarrow}^{\dag} \zeroKet.
\end{eqnarray}
Hence applying the infinitesimal unitary transformation $\infinitesimalU$ with this operator to our example state $\ourExamplePsi$, we obtain a $w$ matrix with extra terms
$\pm \iet$
in the $\aup \bup$ and $\bdown \adown$ locations:
\begin{eqnarray}
	w \rightarrow
	\quarter
	\begin{pmatrix}
		0			& 1				& -\iet		& 1 \\
		-1		& 0				& -1			& -\iet \\
		+\iet	& 1				& 0				& 1 \\
		-1		& +\iet		& -1			& 0
	\end{pmatrix}
\end{eqnarray}
which has a Schliemann entanglement
\begin{eqnarray}
	\eta = \half \epsilon^{2} t^{2} = O(\epsilon^{2}).
\end{eqnarray}

Thus, to first order in $\epsilon$, the Schliemann entanglement of our example state is unaffected: even though the transformation introduces new correlations between the spin states of the particles on sites $A$ and $B$.

%.................................................................%
\paragraph*{Understanding the anomalous behaviour of the Schliemann measure in terms of the Slater decomposition.}

The Slater decomposition representation of an entangled two-fermion, two-site state described earlier provides a particularly simple way of understanding why the Schliemann measure does not behave correctly under either two-site one-particle or one-site two-particle unitary transformations.

According to Schliemann \emph{et al.}, a two-fermion two-site state is entangled iff it has a Slater rank greater than one. It is well-known that a one-particle unitary transformation applied to a Slater determinant will produce another Slater determinant, whereas a two-particle transformation will produce a superposition of Slater determinants. Therefore any one-particle two-site unitary transformation will not affect the Slater rank of a state and so will not change the Schliemann entanglement, despite being a non-local transformation. Similarly, all two-particle one-site unitary transformations will modify the Slater rank of a two-fermion two-site state, and therefore will change the Schliemann entanglement, even though they are local. Schliemann's measure therefore fails to behave as we expect.  The entanglement measures introduced in \cite{paskauskas2001} and \cite{li2001} suffer from analogous problems, since both are based on the rank of the state.

%-----------------------------------------------------------------%
\subsection{Zanardi measure}

Zanardi \cite{zanardi2001} considers the Fock space of $N$ spinless fermions in a lattice with $L$ sites. The state space $H_{L}(N)$ for this system is given by
\begin{eqnarray}
	H_{L}(N) := \text{span} \{ \ket{A} / A \in \mathcal{P}_{L}^{N} \}
\end{eqnarray}
where the antisymmetrized state vector $\ket{A}$ is given by the Slater determinant
\begin{eqnarray}
	\ket{A} := \frac{1}{\sqrt{N!}} \sum_{P \in S_{N}} (-1)^{|P|}  \otimes ^{N} _{l=1} \ket{\psi_{j_{P(l)}}}
\end{eqnarray}
and where $\mathcal{P}_{L}^{N}$ denotes the family of N-site subsets of the site labels, and $\ket{\psi_{j_{P(l)}}}$ is the single particle state for the $j^{\text{th}}$ site where j is a member of the subset $\mathcal{P}_{L}^{N}$.

For some $\PsiKet \in H_{L}(N)$, the local density matrix for the $j^{\text{th}}$ site is given by
\begin{eqnarray}
	\rho_{j} := \text{tr} _{j} \PsiKet \PsiBra
\end{eqnarray}
where $\text{tr} _{j}$ denotes the trace over all but the $j^{\text{th}}$ site, and therefore the von Neumann entropy of $\rho_{j}$ is a measure of the entanglement of the $j^{\text{th}}$ site with the remaining N-1 sites. We will now show that, unlike the other candidates, Zanardi's measure possesses all the desirable features of an entanglement measure that we have listed above.

%=================================================================%
\section{`Site entropy' entanglement measure}
\label{section:siteEntropyMeasure}

%-----------------------------------------------------------------%
\subsection{Application to an example state}

Let us now investigate further the properties of Zanardi's `site entropy' entanglement measure. Our conclusions are that using a binary site-spin occupation number basis for the full density operator for an entangled system, and then reducing the density operator with respect to this basis, gives a reduced density matrix whose von Neumann entropy appears to be a correct measure of entanglement under all circumstances, and for all spin statistics. This is due to the fact that Fock space (to which this representation maps the Hilbert space of a set of indistinguishable particles) has a natural product structure.

For example, for the (fermionic or bosonic) state considered in a previous section
\begin{eqnarray}
	\ourExamplePsi,
\end{eqnarray}
the density operator for the full system is
\begin{eqnarray}
	\rho=\ourExampleRho.
	\nonumber \\
\end{eqnarray}

We now express this density operator as a density matrix in the binary $\{\naup,\nadown,\nbup,\nbdown\}$ occupation number basis, and reduce it for side $B$ by tracing out states of side $A$ using combinations of
$\naup=0,1$  and  $\nadown=0,1$
since the number of particles on site $A$ is
$0, 1\uparrow, 1\downarrow$, or 2. Thus we perform
\begin{eqnarray}
	\rhoB = && \sum_{\naup=0,1, \nadown=0,1}
	\nonumber \\
	&& \aDiagonalElement{\bMatrixElement{\rho}}
	,
	\nonumber \\
\end{eqnarray}
giving
\begin{eqnarray}
	\rhoB = \ourExampleRhoB
\end{eqnarray}
in the \{$\nbup,\nbdown$\} = \{0,0\}, \{1,0\}, \{0,1\}, \{1,1\} basis. The von Neumann entropy of this is
\begin{eqnarray}
	S(\rhoB) = - \text{tr} (\rhoB \  \text{log} _{2} \  \rhoB)
					 = -4 (\quarter \  \text{log} _{2} \quarter)
					 = 2.
\end{eqnarray}

Therefore, according to Zanardi's site entropy entanglement measure our example state contains two ebits of entanglement: one in the spin degree of freedom, the other in the spatial degree of freedom. This will be discussed at length later in this paper. By contrast, as we have seen above, the Schliemann measure gives zero entanglement for this state. We give in Appendix $A$ an explicit construction showing that two qubits may be teleported using this state, further supporting the entanglement value given by the Zanardi measure.

%.................................................................%
\subsection{Behaviour under unitary transformations}

\paragraph*{One-site two-particle (local) unitary transformations}

As before, we apply the infinitesimal one-site, two-particle unitary transformation $\infinitesimalU$ with $H = \oneSiteTwoParticleH$. We obtain
\begin{eqnarray}
	\rhoB = \frac{1}{4}
	\begin{pmatrix}
		1+\epsilon^{2}U^{2} & . & . & . \\
		. 									& 1 & . & . \\
		. 									& . & 1 & . \\
		. 									& . & . & 1
	\end{pmatrix}
	.
\end{eqnarray}
Hence, unlike the Schliemann measure, to first order in $\epsilon$ the site entropy measure is invariant under one-site two-particle unitary transformations. This is the correct behaviour for an entanglement measure: we cannot generate entanglement through a purely local unitary transformation.

\paragraph*{Two-site one-particle (non-local) unitary transformations}

Let us apply the transformation generated by (\ref{eqnModifiedHoppingH}) to our example state $\ourExamplePsi$. Tracing out site $A$, we obtain the reduced density matrix for site $B$,
\begin{eqnarray}
	\rhoB = \quarter
	\begin{pmatrix}
		1	& .								& .								& . \\
		.	& 1								& -2i \epsilon t	& . \\
		.	& +2i \epsilon t	& 1								& . \\
		. & .								& .								& 1
	\end{pmatrix}
\end{eqnarray}
in the $\siteBOccupationNoBasis$ basis. To first order in $\epsilon$ this is not equal to the untransformed $\rhoB$. Therefore, two-site unitary transformations can generate entanglement in the site entropy picture, even if they only operate on one (delocalized) particle. This conclusion is as we would expect.

%-----------------------------------------------------------------%
\subsection{Site entropy measure applied to a completely general state}

%.................................................................%
\paragraph*{Bosonic particles.}

Let us now apply the site entropy description of entanglement to completely general two-particle, two-site states. Since the case of bosonic particles is the most general, we consider it first. The state can now be written in terms of the $w$-matrix as
\begin{eqnarray}
	\schliemannState{b}
\end{eqnarray}
where $1,2,3,4 = \aup,\adown,\bup,\bdown$ and $w$ is now a symmetric coefficient matrix.

Transforming this to the site-spin occupation number basis $\{\naup,\nadown,\nbup,\nbdown\}$ and tracing out the states of site $A$, we obtain a reduced density matrix for site $B$ of block diagonal form, where each block corresponds to a particular occupancy (0,1, or 2 bosons) of that site.
\begin{eqnarray}
	\rhoB^{bosonic} =
	\begin{pmatrix}
		\rhoHat_{B,0}	& .		& . \\
		.		& \rhoHat_{B,1}	& . \\
		.		& .		& \rhoHat_{B,2}
	\end{pmatrix}
\end{eqnarray}
This is a $6 \times 6$ matrix, rather than the $4 \times 4$ $\rhoB$ we previously obtained for the two-fermion state, because Bose-Einstein spin statistics permit the extra site-$B$ double-occupancy states $\bup \bup$ and $\bdown \bdown$.

The `zero particles on site $B$' component is
\begin{eqnarray}
	\rhoHat_{B,0} = |w_{11}|^{2} + |w_{22}|^{2} + 4|w_{12}|^{2}
\end{eqnarray}

The `one particle on site $B$' component in the $\bup,\bdown$ basis is
\begin{eqnarray}
	\rhoHat_{B,1} =
	\begin{pmatrix}
		4|w_{13}|^{2} + 4|w_{23}|^{2} &
		4 w_{13} w_{14}^{*} + 4 w_{23} w_{24}^{*} \\
		
		4 w_{13}^{*} w_{14} + 4 w_{23}^{*} w_{24} &
		4|w_{14}|^{2} + 4|w_{24}|^{2}
	\end{pmatrix}
\end{eqnarray}

Finally, the `two particles on site $B$' component in the \\
$\{ \bup\bdown, \bup\bup, \bdown\bdown \}$ basis is
\begin{eqnarray}
	\rhoHat_{B,2} =
	\begin{pmatrix}	
		4|w_{34}|^{2} &
		2 w_{34} w_{33}^{*} &
		2 w_{34} w_{44}^{*} \\
		
		2 w_{34}^{*} w_{33} &
		|w_{33}|^{2} &
		w_{33} w_{44}^{*} \\
		
		2 w_{34}^{*} w_{44} &
		w_{33}^{*} w_{44} &
		|w_{44}|^{2}
	\end{pmatrix}
\end{eqnarray}

%.................................................................%
\paragraph*{Fermionic particles}

Obtaining an expression for $\rhoB$ for a completely general fermionic state is simply a matter of applying the Pauli exclusion principle to $\rhoB^{bosonic}$. Under Fermi-Dirac statistics, the only possible two-particle state on site $B$ is $\bup \bdown$, meaning that the two-particle part of $\rhoB$ is reduced to the $1 \times 1$ submatrix
\begin{eqnarray}
	(\rhoHat_{2,B})_{\text{fermionic}} = (4|w_{34}|^{2})
\end{eqnarray}

Similarly, the only possible two-particle state on site $A$ is $\aup \adown$, meaning that the probability of zero particles on site $B$ is given by $4|w_{12}|^{2}$. Hence the zero-particle part of $\rhoB^{bosonic}$ is
\begin{eqnarray}
	(\rhoHat_{0,B})_{\text{fermionic}} = (4|w_{12}|^{2})
\end{eqnarray}

The one-particle part of $\rhoB^{bosonic}$ is by definition not affected by Pauli exclusion, therefore
\begin{eqnarray}
	(\rhoHat_{1,B})_{\text{fermionic}} = (\rhoHat_{1,B})_{\text{bosonic}}
\end{eqnarray}

%-----------------------------------------------------------------%
% Source is `OccupationNoEntanglementNotes.nb'

\subsection{Relationship to Wootters tangle}

The origin of the Wootters entanglement measure is now readily understood. It is simply the von Neumann entropy of the one-particle part $\rho_{1}^{B}$ of the reduced density matrix in the occupation number representation for site $B$. Wootters's `general state' equation~(\ref{eqnWoottersGeneralState}),
where the kets represent
$\ket{\sigma_{A} \  \sigma_{B}}$,
can be rewritten in the occupation number basis~$\occupationNoBasisForWoottersGeneralState$ as
\begin{eqnarray}
	\woottersGeneralStateInOccupationNoBasis.
\end{eqnarray}

Tracing out site $A$ yields the following reduced, correctly normalized, density matrix for site $B$ in the $\bup, \bdown$ basis:
\begin{eqnarray}
	\woottersRhoBInOccupationNoBasis
\end{eqnarray}
with eigenvalues
\begin{eqnarray}
	\woottersRhoBUnsimplifiedEigenvalues.
\end{eqnarray}
Applying the simplifications $\woottersTauDefn$ and $\woottersXDefn$ these reduce to
$1-x,x$. Thus the entropy of $\rhoB$ is
\begin{eqnarray}
	S(\rhoB) = -(x \text{log} _{2} x + (1-x) \text{log} _{2}(1-x))
\end{eqnarray}
which is identical to the Wootters result for entanglement given in (\ref{eqnWoottersE}).

%=================================================================%
\section{Spin-space entanglement transfer}
\label{section:entanglementTransfer}

%-----------------------------------------------------------------%
\subsection{Omar \emph{et al.} thought experiment}
\label{omarSubsection}

% Source for this is `OmarInOurFormalism.nb'

\begin{figure*}
	\caption
	{
		\label{fig:omar_apparatus}
		The spin-space entanglement apparatus used in the Omar et al. thought experiment (reproduced from \cite{omar2001}).
	}

	\includegraphics{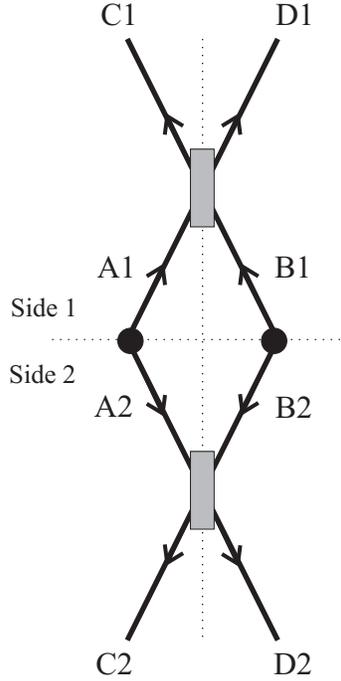}
\end{figure*}

Since we have argued that Zanardi's approach gives a correct view of entanglement in all circumstances, we can use it to analyze situations in which there is spatial, as well as spin, entanglement. A particularly interesting system of this type was introduced recently by Omar \emph{et al.} \cite{omar2001}. They consider an apparatus which takes as its input two pairs of particles, $A$ and $B$, each pair maximally entangled in some internal degree of freedom (e.g. spin), and transfers some of that entanglement to the spatial degrees of freedom of the particles. This is achieved by passing one particle from each pair through a beam splitter on one side of the apparatus, and doing likewise with the remaining particles from each pair through another beam splitter on the other side of the apparatus (see Figure \ref{fig:omar_apparatus}). The two sides are labelled 1 and 2. Use of the site entropy measure enables us to understand better the process of entanglement transfer.

%.................................................................%
\paragraph*{Side 1 of the apparatus.}

First, let us consider the input state to the apparatus, and its entanglement according to the site entropy measure. This state is
\begin{eqnarray}
	\omarPlusPlusInputState
	.
\end{eqnarray}
Henceforth we will consider the case where all four particles are fermions, and the above product state consists of triplets (described as the `$++$ case for fermions' in \cite{omar2001}). If we write this in the occupation number representation, and then trace out side 2 of the apparatus, we obtain the following reduced density matrix for side 1 of the apparatus:
\begin{eqnarray}
	\label{eqn:omarInputRhoOne}
	\omarInputRhoOne
\end{eqnarray}
using the reduced basis for side 1
\begin{eqnarray}
	\omarRhoOneBasis
	.
\end{eqnarray}
This state has two ebits of entanglement. Examining equation (\ref{eqn:omarInputRhoOne}), we see that this entanglement is carried entirely in the bottom right part of the density matrix, which corresponds to single-occupancy states which differ only by the spin. Therefore, this entanglement is purely spin entanglement.

It is a straightforward exercise to show that the site entropy measure gives the same total entanglement between sides 1 and 2 of the apparatus (two ebits) for the input and output states. This must be so since the operation of each beamsplitter is local to its side of the apparatus. The unnormalized output state for 50/50 beam-splitters for our input state is given in \cite{omar2001} as
\begin{eqnarray}
	\omarFermionOutputState
\end{eqnarray}
where for example, $\LKet_{1}$ indicates both fermions on side 1 of the apparatus have passed into the left arm and thus necessarily have opposite spins, and $\ket{\aupdown}_{1}$ indicates that each particle on side 1 of the apparatus has passed into a different arm, with the particle occupying the left arm being spin up, the particle occupying the right arm being spin down.

If we rewrite this in the occupation number representation
\begin{eqnarray}
	\occupationNoRepresentationForOmar ,
\end{eqnarray}
trace out side 2 of the apparatus, and renormalize, we obtain the following reduced density matrix for side 1 of the apparatus:
\begin{eqnarray}
	\label{eqn:omarOutputRhoOne}
	\omarOutputRhoOne
\end{eqnarray}
which has entropy $S(\rhoHat_{\text{1,out}}) = 2$, showing that the total entanglement is unaffected by the operation of the apparatus. However, we can see from the fact that the double-occupancy top-left sector of this matrix is now non-zero that the system now contains spatial entanglement, because this state is now mixed in arm-occupancy number as well as spin. 

%%%% Comparing equations (\ref{eqn:omarInputRhoOne}) and (\ref{eqn:omarOutputRhoOne}) we can see that
%%%% non-zero double-occupancy diagonal elements of the reduced density matrix correspond to non-zero spatial entanglement.

%.................................................................%
\paragraph*{Single-occupancy and double-occupancy entanglements are additive.}

Since in (\ref{eqn:omarOutputRhoOne}) there are no non-zero off-diagonal elements connecting the double-occupancy and single-occupancy sectors of the matrix, we can unambigously assign each eigenvalue to one sector, and hence divide the total entanglement into double-occupancy and single-occupancy parts. In this case, the double-occupancy sector has eigenvalues $\quarter,0$ and hence contributes $0.5$ ebits to the entanglement. The single-occupancy sector has eigenvalues of $\quarter,0,\quarter,\quarter$ and hence contributes $1.5$ ebits. It is clear from these definitions that the single-occupancy and double-occupancy entanglements will always sum to the total entanglement, provided that the off-diagonal elements connecting the two sectors are zero. The single-occupancy entanglement is a form of spin entanglement, since the single-occupancy states do not differ in the spatial distribution of particles between the arms. Likewise, the double-occupancy entanglement is a form of space entanglement, since the double-occupancy states do not differ in their $m_{s}$ values. However it is not obvious that these are the most general forms of spin and space entanglement, since for example, the double-occupancy entanglement does not take account of the spatial states in which each arm contains one particle.

The distinction between spatial and double-occupancy entanglement is further clarified by the $S_{x}=0$ spin measurements suggested by Omar et \emph{al.} for their output state. They show that the spatial state produced by such a measurement (obtained with probability $\half$) involves a superposition of both double- and single-occupancy components. In this state, they show that the entanglement remaining between sides $1$ and $2$ is one ebit: since the spin state is now the same for all components and hence unentangled, this could be unambiguously described as spatial entanglement.

%.................................................................%
\paragraph*{Left arm of side 1 of the apparatus.}

It is instructive now to reduce further the input and output density matrices to those for just the left arm of side 1 of the apparatus. For the input state, this is
\begin{eqnarray}
	\label{eqn:omarInputRhoOneLeft}
	\omarInputRhoOneLeft
\end{eqnarray}
in the basis
\begin{eqnarray}
	\omarRhoOneLeftBasis
\end{eqnarray}
which has entropy $S(\rhoHat_{\text{1L,in}}) = 1$.
In the same basis, the reduced density matrix for the output state is
\begin{eqnarray}
	\label{eqn:omarOutputRhoOneLeft}
	\omarOutputRhoOneLeft
\end{eqnarray}
which has entropy $S(\rhoHat_{\text{1L,out}}) = 1.81$, showing that the action of the beamsplitter on side 1 of the apparatus has introduced an additional 0.81 ebits of entanglement between the left arm of side 1 and the rest of the system, in addition to the 1 ebit of entanglement already present between those two subsystems.

%.................................................................%
\paragraph*{Operator-sum representation for spin-space entanglement transfer.}

It is possible to find an operator-sum representation for the spin-space entanglement transfer within the left arm of side 1 of the apparatus that we have discussed above. An easy way to do this is to make the following isomorphism between the spin states of two qubits A and B, and the occupation numbers for the spin-up and spin-down single particle states of the left arm:
\begin{eqnarray}
	\label{eqn:twoQubitSpinStatesToTwoParticleOccupationNosMapping}
	\twoQubitSpinStatesToTwoParticleOccupationNosMapping
	.
\end{eqnarray}
We then find that the action of the Omar interferometer in transforming $\rhoOneLeftIn$ to $\rhoOneLeftOut$ can be represented by the action of the depolarizing channel
\cite{preskillBook}
on $\rhoOneLeftIn$ with probability $p=\frac{3}{8}$.

%-----------------------------------------------------------------%
\subsection{Division of entanglement into single- and double-occupancy parts}

When is it possible to divide entanglement unambiguously into single- and double-occupancy parts? As we can see from the above treatment of the Omar apparatus, it is when the reduced density matrix for the subsystem whose entanglement we are considering has sectors corresponding to single- and double-occupancy, with no off-diagonal elements connecting them. Such a division is possible whenever the total system contains a definite number of particles: there are then no off-diagonal density matrix elements connecting states of the subsystem having different numbers of particles. This is the reason why there are no elements connecting the different sectors of $\rhoHat_{1L}$ in equations (\ref{eqn:omarInputRhoOneLeft}) and (\ref{eqn:omarOutputRhoOneLeft}).

We note with emphasis that the situation for the reduced density matrix for side 1 of the Omar apparatus in equation (\ref{eqn:omarOutputRhoOne}) is fundamentally different. All the basis states contain the same total number of particles. Its block-diagonal form is due to a combination of factors: the spin symmetry of the system (which causes those elements connecting states on side 1 and 2 with a different total $m_{s}$ value to be zero), and the use of 50/50 beam splitters, which prevents any products of the form $\LKet_{1} \AKet_{2}$ from appearing in the output state.

%=================================================================%
% All sections after here are appendices.

\appendix

%=================================================================%
\section{Teleporting two qubits using an example delocalized state}

%.................................................................%
\paragraph*{Protocol design.}

Consider again the delocalized state in equation (\ref{eqn:ourExamplePsi}). Since the Zanardi measure says it contains two ebits of entanglement, we should be able to teleport two qubits of quantum information using it. Clearly, since the two ebits are spread across spin and spatial degrees of freedom, we shall need to modify the original protocol somewhat. How could we do this?

Switching into the anthropocentric language of `Alice' and `Bob', a concise description of the protocol for teleporting one qubit described in \cite{bennett1993} is as follows. We separate the two subsystems of our entangled system which will act as the channel for quantum information, giving one to Alice and one to Bob. We then perform a CNOT on the qubit whose state we wish to teleport (the `source qubit'), and Alice's system, using the source qubit as the control line. We then perform a Hadamard transform on the source qubit. Alice's qubits are now in a superposition of states, each of which corresonds to the target qubit being in the same state as the original state of the source qubit up to a unitary transform. Alice performs a measurement of the state of her two qubits, thereby projecting the target qubit into one of these states. The protocol is completed by Alice sending Bob two bits of classical information describing which measurement result she obtained, enabling him to rotate the target qubit into the correct state.

The key to teleporting via the delocalized state (\ref{eqn:ourExamplePsi}) lies in recognizing that the two ebits in the delocalized state are equivalent to two pairs of qubits, each of which is maximally spin-entangled (`channel pairs'), and making the following isomorphism:
\begin{eqnarray}
	\label{eqn:isomorphismForTwoQubitTeleportation}
	\isomorphismForTwoQubitTeleportation
\end{eqnarray}
This connects the occupation numbers of the single particle states of Alice's site to the states of Alice's channel-pair qubits in the spin-only representation.

Recall that a CNOT performs
\begin{eqnarray}
	\spinOnlyCNOTDefinition
\end{eqnarray}
i.e. we flip the second qubit in a basis state iff the state of the first (control) qubit in that basis state is `up'. What does a CNOT on \textbf{one} of Alice's two channel-pair qubits look like after applying the above isomorphism? Using this basis for the states of one of the source qubits ($C$) and Alice's site ($A$):
\begin{eqnarray}
	\aliceTeleportationOccupationNoBasis
\end{eqnarray}
we obtain the following unitary transformation for the first `virtual' qubit:
\begin{eqnarray}
	\label{eqn:cnotOnVirtualFirstQubitInDelocalizedState}
	\cnotOnVirtualFirstQubitInDelocalizedState
	.
	\nonumber \\   % NB. This linebreak prevents the matrix running into the text below!
\end{eqnarray}

The action of this is thus:
\begin{eqnarray}
	\actionOfVirtualFirstQubitCNOT
	.
\end{eqnarray}
Referring to the isomorphism in (\ref{eqn:isomorphismForTwoQubitTeleportation}) we see that this $\hat{U}$ flips the first `virtual' qubit in Alice's half of the delocalized state iff the control qubit $C$ is spin-up. Similar considerations lead to a similar unitary transformation for the second `virtual' qubit. We also note that since we are teleporting two qubits, we need to send four classical bits to complete the protocol.

%.................................................................%
\paragraph*{Protocol implementation.}

The two CNOTs described above will clearly allow us to exploit the two ebits of entanglement present in the delocalized state. However some consideration needs to be paid to how we can implement these CNOTs. Considering again the first `virtual' qubit, the Hamiltonian we can use to generate equation (\ref{eqn:cnotOnVirtualFirstQubitInDelocalizedState}) is
\begin{eqnarray}
	\hamiltonianForCNOTOnVirtualFirstQubitInDelocalizedState
	.
\end{eqnarray}
In the first expression we have used the bases
\begin{eqnarray}
	\basesForHamiltonianForCNOTOnVirtualFirstQubitInDelocalizedState
	.
\end{eqnarray}
In the second expression we have rexpressed the projectors for the occupation number state of site $A$ in second-quantized notation. Similarly, the Hamiltonian generating a CNOT on the second `virtual' qubit is
\begin{eqnarray}
	\hamiltonianForCNOTOnVirtualSecondQubitInDelocalizedState
	.
\end{eqnarray}

Neither of these Hamiltonians conserves particle number, thus we need to introduce a coherent source/sink of particles to the system. We shall see in a moment that we can easily do this for bosons. Introducing a system $D$ which acts as a particle source/sink,
$\hat{H}_{\text{1st qubit}}$
becomes
\begin{eqnarray}
	\hamiltonianForCNOTOnVirtualFirstQubitInDelocalizedStateWithParticleSourceSink
	.
\end{eqnarray}

At this point we face a problem. By changing the number of particles in system $D$ as a consequence of our CNOT, we are introducing new correlations between the states of subsystems $A$ and $D$. This is thus a type of decoherence affecting the entanglement of the `carrier-pair' $AB$. This is clearly unavoidable in a real-world system, but we can show that for bosons, by choosing a suitable initial state for subsystem $D$ we can minimize this decoherence to a negligible level. We seek to put system $D$ in an approximate eigenstate of the creation and annihilation operators, so that they leave it unchanged and no decoherence of the entanglement in the $AB$ carrier pair occurs. A suitable choice is the coherent state
\begin{eqnarray}
	\coherentStateDefinition
	.
\end{eqnarray}
It is well known that this state is an eigenstate of the annihilation operator, a fact which suits our requirements perfectly, but it is not an eigenstate of the creation operator. However, as the mean number of particles $|\alpha|^{2}$ in the coherent state asymptotically approaches $\infty$, the state asymptotically approaches an eigenstate of the creation operator. It is important to note that this method for coherently producing a non-number-conserving interaction applies to bosons only. For fermions, Pauli exclusion prevents us using such a simple approach and there is no analogue of the coherent state available within the Hilbert space.

%-----------------------------------------------------------------%
% Source for this is `OrthogonalizeFermionBellStates.nb'

%-----------------------------------------------------------------%
% Reproduced from Omar et al with kind permission

%-----------------------------------------------------------------%
\bibliographystyle{unsrt}
\bibliography{qph}

\end{document}